# CODING BY DESIGN: GPT-4 EMPOWERS AGILE MODEL DRIVEN DEVELOPMENT

Ahmed R. Sadik[1][a], Sebastian Brulin[1][b] and Markus Olhofer[1][c]
[1]*Honda Research Institute Europe, Carl-Legien-Strasse 30, Offenbach am Main, Germany*
*{ahmed.sadik, sebastian.brulin, markus.olhofer}@honda-ri.de*

Abstract: Generating code from a natural language using Large Language Models (LLMs) such as ChatGPT, seems groundbreaking. Yet, with more extensive use, it's evident that this approach has its own limitations. The inherent ambiguity of natural language presents challenges for complex software designs. Accordingly, our research offers an Agile Model-Driven Development (MDD) approach that enhances code auto-generation using OpenAI's GPT-4. Our work emphasizes "Agility" as a significant contribution to the current MDD method, particularly when the model undergoes changes or needs deployment in a different programming language. Thus, we present a case-study showcasing a multi-agent simulation system of an Unmanned Vehicle Fleet. In the first and second layer of our approach, we constructed a textual representation of the case-study using Unified Model Language (UML) diagrams. In the next layer, we introduced two sets of constraints that minimize model ambiguity. Object Constraints Language (OCL) is applied to fine-tune the code constructions details, while FIPA ontology is used to shape communication semantics and protocols. Ultimately, leveraging GPT-4, our last layer auto-generates code in both Java and Python. The Java code is deployed within the JADE framework, while the Python code is deployed in PADE framework. Concluding our research, we engaged in a comprehensive evaluation of the generated code. From a behavioural standpoint, the auto-generated code aligned perfectly with the expected UML sequence diagram. Structurally, we compared the complexity of code derived from UML diagrams constrained solely by OCL to that influenced by both OCL and FIPA-ontology. Results indicate that ontology-constrained model produce inherently more intricate code, but it remains manageable and low-risk for further testing and maintenance.

[a]  https://orcid.org/ 0000-0001-8291-2211
[b]  https://orcid.org/ 0000-0002-9710-6877
[c]  https://orcid.org/0000-0002-3062-3829

# 1 INTRODUCTION

In the AI era, with Large Language Models (LLMs) trained on diverse code, new opportunities arise for innovation in Model-Driven Development (MDD). MDD is an evolving field that holds promise to improve the efficiency and robustness of software engineering practices (Hailpern & Tarr, 2006). This study introduces an agile MDD approach that leverages the use of existing LLMs such as OpenAI's ChatGPT, to auto-generate complete, deployment-ready software artifacts (Sadik, Ceravola, et al., 2023). Our approach eliminates the intensive time and effort seen in conventional MDD, where it is needed to craft a unique code generator for each deployment or update the generator with every model alteration. Complete software implies that the auto-generated artifacts are not only intricate but also synergistically structured to collectively ensure their designated functionality and meet its specified requirements (Feltus et al., 2017).

Code generation, especially from formal models such as Unified Modeling Language (UML), Systems Modeling Language (SysML), or Business Process Model and Notation (BPMN) diagrams has emerged as an influential paradigm in modern software engineering practices (OMG, 2006). Class diagrams, a primary component of most object-oriented design methodologies, capture the static structure of software systems by representing classes, their attributes, operations, and their interrelationships. When paired with the appropriate tools, these diagrams can be directly converted into executable code, facilitating a more streamlined software development process (Sharaf et al., 2019). Such automation not only guarantees a solid alignment between the design and its corresponding implementation but also reduces manual coding errors. This leads to improved software quality and quicker market deployment (Sarkisian et al., 2022).

The dynamic behaviour, interactions, and holistic views of the system play a fundamental role in comprehending its overall functionality. UML offers a suite of diagrams, each with a unique perspective on system modelling. For instance, UML sequence diagrams represents the interactions among objects in a time-sequential manner, capturing the intricacies of object communications(Perez-Martinez & Sierra-Alonso, 2004). Use case diagrams focus on the system's functionalities from an end-user's perspective, ensuring the system's relevance and usability. Moreover, state diagrams offer insights into the various states an object can have and the triggering events for state transitions. When code generation processes solely rely on the static structure offered by class diagrams, they might miss out on these dynamic and interactional aspects of system behaviour. Thus, for truly comprehensive and complete auto-generated code, there is an imperative need to synergistically combine the static semantic richness with the dynamic perspectives offered by other UML diagrams (Kapferer & Zimmermann, 2020).

However, while class diagrams focus on the structural aspects of a system, they frequently miss capturing intricate rules, constraints, or the specifications inherent to a domain. This is where the Object Constraint Language (OCL) plays a pivotal role in adding the code construction fine details, as it offers a declarative language to specify precise constraints and derived values, that are vital for maintaining the integrity and consistency of the model (Cabot & Gogolla, 2012). Furthermore, the dynamic constraints of the model such as communication between the classes can be constrained as well via a domain-specific ontology language such as FIPA-ontology (FIPA, 2000). Domain-specific ontology enables the common understanding of knowledge that is exchanged via bridging the semantic gap that conventional class diagrams do not include (Siricharoen, 2009). The integration of class diagrams, combined with OCL constraints and domain-specific ontologies, could spearhead a novel epoch in code generation. This integrated approach would enable the production of code that's not merely structurally accurate but also enriched with semantic details, ensuring the resulting software mirrors both its foundational design and the detailed domain expertise.

Evaluation of auto-generated code is an essential step in grasping its software quality (Liu et al., 2023). Traditionally, criteria to assess this quality, such as testability, maintainability, and reliability, have been qualitative in nature. This inherent qualitative character has often rendered them relative and open to subjective interpretation. Accordingly, our research tends to adopt more objective, quantifiable criteria. Given our primary objective to auto-generate seamless code from a model, our evaluation lens sharply focused on the structural integrity of this auto-generated code. We utilized cyclomatic complexity as an instrumental metric to provide insights into its structural soundness. Furthermore, a comparative analysis was executed, pitting the behaviors of the generated code—deployed in varied languages—against each other and against the expected behavior from a model's perspective (Ahmad et al., 2023).

The paper is structured to guide the reader through our study. Section 2 provides a detailed problem statement, where we pinpoint the current challenge with existing MDD approach that hindering it from become agile. Section 3 breaks down the four layers of the proposed agile MDD approach. Section 4 applies the proposed approach to model a Multi-Agent System (MAS) of an Unmanned Vehicle Fleet (UVF) and Mission Control Center (MCC). After modeling the structure of the case-study using a comprehensive UML class diagram, we add two layers of constraints to constrain both the model construction and communication. We used OCL to describe the model details such as invariants, pre, and post conditions. Furthermore, we used Intelligent Physical Agents (FIPA) ontology, to define the communication semantics among the agents. Then we modelled the case-study behaviour by UML use case diagram, activity diagram and state machine. Finally, we auto-generate Java and Python code from the same model by using GPT-4. The code is deployed to simulate the UVF-MCC multi agent in The Java code is deployed within the Java Agent Development (JADE) framework, and Python Agent Development (PADE) framework consequently. In Section 5, we evaluate the model behavior by comparing JADE and PADE simulation behavior to each other's and to the originally designated predicated behavior from the model. Furthermore, we took a closer look at the generated autocade structure via its cyclomatic complexity. In this part we compared a generated code complexity from a model with only OCL constrains to the same model that is constrained by OCL and FIPA-ontology. Finally, Section 6 wraps up our findings, discusses their implications, and suggests next steps in future research.

## 2 PROBLEM STATEMENT

Natural language inherently possesses ambiguity, which is not only a challenge for machines to comprehend but is also confusing for humans. When utilizing ChatGPT to auto-generate intricate software artifices, defected code is often produced, due to the uncertain and open-ended nature of the input prompt. This issue becomes significantly pronounced in cases where the software to be generated is complex, multi-dimensional and cannot be effectively described using natural language.

Yet, MDD promises an elevated level of software abstraction, where high-level model is used as the primary artifact from which the final application is generated (Sadik & Goerick, 2021). However, the process of designing and maintaining these models, introduces challenges that can limit the MDD effectiveness. Herein, two main problems can be identified. Firstly, traditional modeling techniques, such as UML diagrams, while being excellent for data structuring in software development, often lacks the semantic richness and rule-based derivation of knowledge inherent to ontologies (Belghiat & Bourahla, 2012). This leads to models that are accurate in terms of structure and behavior but lacking in semantic depth, making them less effective in modeling complex real-world scenarios. Secondly, transforming these models into executable code is not a straightforward process (Petrovic & Al-Azzoni, 2023). As it involves manual scripting of the code generator, that requires to be meticulously maintained and updated to keep pace with changes in the model and the underlying technology stack. This is particularly true when alternating the deployment from one programming language to another (Cámara et al., 2023)

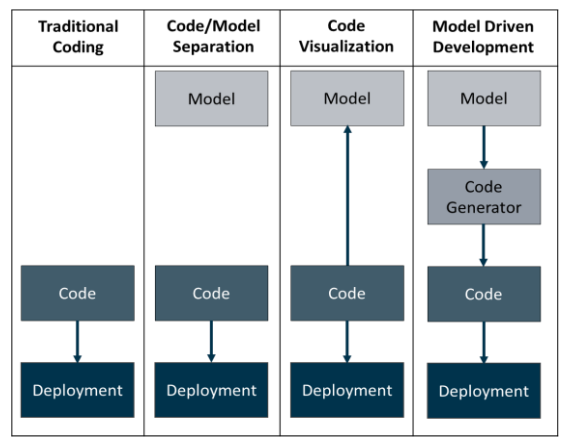

Figure 1: Difference between traditional coding and MDD.

To comprehensively address the challenge in code generation within the framework of the current MDD approach (Kelly & Tolvanen, 2008), it is essential to pinpoint the distinction between traditional coding and MDD code generation as shown in Figure 1. Traditional coding tends to directly encode the software functionalities in code. This approach works well for smaller features that can be transcribed straightforwardly as code. Debugging, testing, and maintenance are also performed at the code level. In contrast, the model-and-code separation approach involves the use of models to abstract and understand the system better, separate from the code. Developers interpret these models while coding the application, and once coding is done, models are often discarded due to the high cost of keeping them up to date. Code visualization involves creating models after the software is designed and built, to understand what a program does or to import libraries or other constructs from code to be used as elements in models. These models, however, are typically not used for implementing, debugging, or testing the software as we have the code (Fadjukoff & Tolvanen, 2022).

While in MDD, models are the primary artifacts in the development process. These source models are used instead of source code. The target code is automatically generated from these models, which raises the level of abstraction and hides complexity. Tools like Eclipse Papyrus, MagicDraw, Enterprise Architect, or IBM Rational Rhapsody have traditionally been used for this purpose (David et al., 2023). Yet, every time there is a shift in the deployment language or a significant update in the model, these tools necessitate substantial alterations to the code generators, impeding agility in the development cycle. However, creating a code generator is a demanding task, consuming considerable time and energy from modelers. Further complicating matters is the requirement to craft a unique code generator for each programming language, making the prevailing MDD approach less adaptive to different deployment languages, and therefore agile MDD fails to exist. It's in this context that we see potential in leveraging LLMs like ChatGPT as universal code generators (Chen et al., 2021). Accordingly, our study highlights the challenging issue that "Although, MDD provides a structured methodology that overcomes the inadequacies of natural language for auto-generating deployable code, the existing MDD approach has not been adequately adapted to the present LLM capabilities in code auto-generation. This misalignment makes the current MDD approach unfit for the agile software development workflow."

## 3 PROPOSED APPROACH

To tackle the challenge outlined in the problem statement, our proposed MDD approach necessitates that ChatGPT fully understands the model and its associated views. Given that ChatGPT currently processes information through text prompts, we employed PlantUML to convert the visual UML diagram to formal text representation, that can be easily copied into ChatGPT prompt.

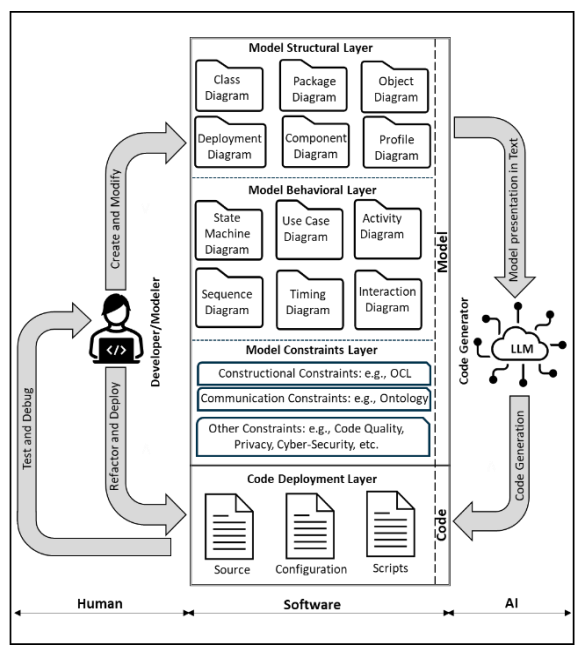

Figure 2: Proposed Agile Model Driven Development approach.

In the proposed approach in Figure 2. The modeller initially creates the different model layers, which are structural, behavioural, and constraints. The structural layers contain all the diagrams that reflect the static by illustrating the software components and the intricate relationships among them. For example, Class diagrams detail object relationships and hierarchies, while package diagrams group these objects, highlighting dependencies. Component diagrams then break down system functionality at a high level, capturing inter-component relationships. For the real-world physical layout, deployment diagrams depict hardware configurations and component distributions. Object diagrams offer runtime object snapshots, while profile diagrams tailor UML models to specific platforms. The behavioural layer models how the system operates and interacts. Sequence diagrams lay out events in a linear progression, giving a clear timeline of interactions. Activity diagrams present a flowchart-like representation of processes, detailing step-by-step actions. Interaction diagrams showcase the interplay between components, while timing diagrams emphasize the importance of timing and sequence. On the user side, use case diagrams illustrate how external entities engage with the system. Lastly, state diagrams capture the life cycle of entities, showing how they transition between different states.

Although, the structural and behavioural diagrams provide a holistic architectural view, they often lack the rules that regulate the model architecture semantics. Accordingly, in this research we propose the constraints layer that fine-tunes the model, by explicitly the model meta-values that cannot expressed by UML notations. OCL for example is used to restrict the construction details of the structural and behavioural layers, by specifying invariants on classes and stereotypes, describe pre- and post-conditions on method and states, and limits the parameters' values. Furthermore, communication constraints can be defined using the proper formal method such as an ontology language to express the communication semantics meanings, and protocol that is necessary to communicate and share knowledge among the software artifacts.

Ultimltiy in the code deployment layer, we employed ChatGPT, which is based on the GPT-4 architecture, to generate code. The reason to use GPT-4 rather than GPT-3.5 is the higher capability of GPT-4 to reason, which is very important value for our approach as the LLM must understand the model semantics and rules that is encapsulated in the constraints layer to embed them in the generated code. Furthermore, after the using ChatGPT to auto-generate the code, it is important that the modeler deploys the generated code onto the software platform and ensure that it is operational. However, it's important to be aware that ChatGPT's code generation capabilities are still evolving and not flawless (Dong et al., 2023). As such, it is anticipated that there may be bugs encountered during the deployment of the code. Consequently, it is necessary for the mod to address these bugs, potentially with the assistance of ChatGPT, and repeatedly run the code until it is successfully fulfilling its intended purpose.

Ultimately, in the code deployment layer, ChatGPT, based on the GPT-4 architecture, is employed to generate code. The choice of GPT-4 over GPT-3.5 is due to its superior reasoning capabilities, a vital feature for our approach. The LLM must comprehend the model semantics and semantics encapsulated in the constraints layer to integrate them into the generated code. Furthermore, post auto-generation of code using ChatGPT, it is crucial for the modeler to deploy the generated code and verify its operationality. However, it's essential to acknowledge that ChatGPT's code generation capabilities are continually evolving and may not be flawless. Therefore, the possibility of encountering bugs during the code deployment is anticipated. It becomes imperative for the modeler to address these bugs, possibly with assistance from ChatGPT, and iteratively run the code until it successfully accomplishes its intended purpose."

## 4 CASE-STUDY MODEL

The chosen use-case involves a UVF, comprising various types of UVs that undertake specific missions and are coordinated by an MCC involving a human operator (Sadik, Bolder, et al., 2023). This case-study is intentionally distributed, enabling it to be modeled and simulated as a MAS (Brulin & Olhofer, 2023). Such MAS often encompasses a high complexity level, as each entity is represented as an agent and must communicate and share information with other entities (i.e., agents) to achieve a common goal (i.e., the fleet mission). To avoid overwhelming the reader with the intricacies of MAS in the following sections, we will highlight only the essential model views that facilitate an understanding of the MAS operation concept (Sadik & Urban, 2018).

### 4.1 Model Structural Layer

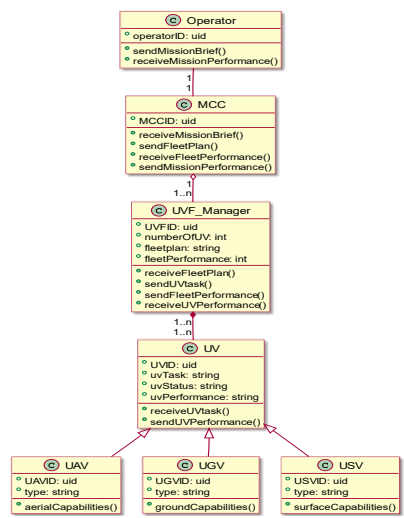

Figure 3: Case-study class diagram.

Class diagram can be considered the most important view in the model structure layer. Every entity within the case-study is represented as an agent class as shown in Figure 3. A summery of these agents are explained as follows:
- **Operator**: models the human operator and contains attributes such as operator-ID. It also includes the actions such as send the mission-brief and receive the mission-performance.

- **MCC**: models the command center, including attributes like MCC-ID. MCC coordinates missions and monitoring the fleet. It includes actions such as receive the mission-brief, send the fleet-plan, receive the fleet-performance, and send the mission-performance.
- **UVF-Manager**: models the UVF-manager, containing attributes like UVF-ID, UVs' number, fleet-plan, and fleet-performance. It includes actions such as receive the fleet-plan, send UV-tasks, send the fleet-performance, and receive the UV-performance.
- **UV**: a generic class that models the UVs. It contains attributes such as UV-ID, the UV's Task, the UV's Status, and Performance. Actions include receive the UV-task and send the UV-performance.
- **UAV, UGV, USV**: these are subclasses of UV, each modelling different types of a UV.

The previously described class diagram has been employed to articulate the intricate internal details of each agent, encompassing attributes, operations, and visibility. Additionally, the class diagram is utilized to delineate all conceivable relationships among the agents, including composition, aggregation, and inheritance. Moreover, the multiplicity of the classes establishes the cardinality between the agents.

## 4.2 Model Behavioural Layer

To maintain brevity and keep the article focused, the article will only explain two views which are the activity and state diagrams, as they are the most important behavioral views to understand the case-study model. The activity diagram refines and complements the class diagram by meticulously detailing aspects such as synchronization, parallel execution, and conditional flows, which are indispensable for effectively achieving the mission goals. In contrast, the state diagram offers a microscopic perspective, unveiling the life cycle of the agents' class within the model and illuminating how they coordinate and respond to realize the overarching mission objectives.

The activity diagram in Figure 4 elaborates the interplay of information and task flows within the agents. It illustrates the orchestration of processes and the sequence in which tasks are allocated, carried out, and assessed, providing an understanding of the MAS temporal and logical dynamics. Thus, the interaction begins when the operator agent sends the mission-brief to the MCC agent. The latter transforms the brief into a plan and conveys it to the UVF-manager, which, in turn, assigns the tasks to the available UVs. Subsequently, the UVF-manager agent awaits the completion of tasks by each UV and collates their performance, which is instrumental in assessing the overall UVF performance. This consolidated performance is relayed to the MCC, translated into mission-performance, and communicated back to the operator agent.

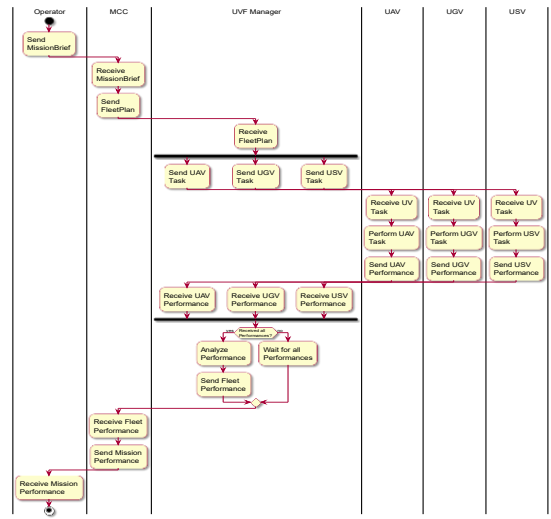

Figure 4: Case-study activity diagram.

Furthermore, the state diagram provides a granular exploration into the agent internal behaviour, illustrating the transitions triggered by events and the corresponding actions undertaken by the agents. In the scenario presented, the operator agent, the MCC, and the UVF-manager are all modelled using a straightforward two-state diagram, representing states of being either busy or free. However, for the UV, modelling a more intricate state machine was imperative, as it is utilized later by the agent to assess its task performance. Figure 5 explains the different UV states as follows:

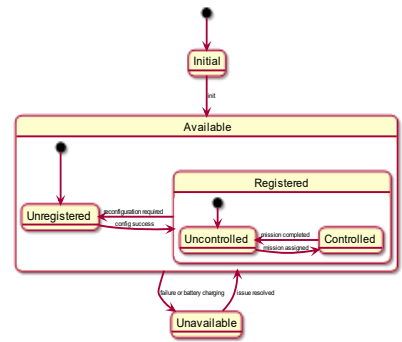

Figure 5: UV agent state diagram.

- **Initial**: the UV is prepared and ready to operate.
- **Available**: the UV can be either registered or unregistered.
- **Unavailable**: the UV is rendered unregistrable due to being out of service, possibly due to a failure or battery charging.
- **Unregistered**: In this condition, the UV is available but has not been registered, as it is still configuring its parameters.
- **Registered**: the UV can be either controlled or uncontrolled.
- **Uncontrolled**: the UV is registered but has not been assigned any mission.
- **Controlled**: the UV is not only registered but also allocated a mission.

## 4.3 Model Constraints Layer

The constraints layer in the proposed MDD approach acts as a meta-model that encapsulates all aspects of the technical requirements that cannot be formalized in the structural and behavioral layers. In the following sections we will discuss in detail the different types of meta-model constraints that have been considered within the case-study modeling.

### 4.3.1 Construction Constraints

The OCL is a declarative language used primarily with UML to describe rules that apply to classes within a model. Incorporating OCL as construction constraints within UML diagrams serves as a key enabler for refining the model. Therefore, OCL enhances the model clarity by addressing the inherent ambiguities in its construction details. This precision is particularly beneficial in the generation of deployed implemented code directly from model views. The added constraints ensure that the transition from model to code is more accurate and seamless.

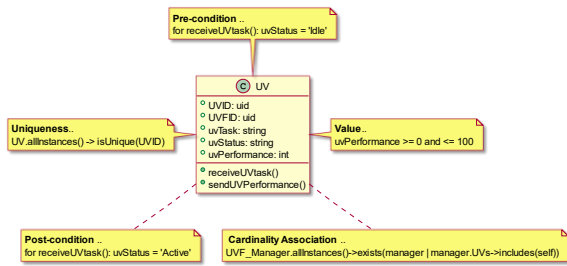

Figure 6: UV agent construction constraints in OCL.

During our study, we categorized five different types on constraints, and we applied them on all the agent classes. Figure 6 shows an example of some of the constrains that applied on the UV agent class. The five constraint types are summarised as follows:

- **Uniqueness**: to ensure that every instance agent is unique. For example, the UV agent must have a unique identifier across MAS.
- **Cardinality**: to ensure the association of the agent instances with each other's, for each UVF-manager with a unique ID is associated with a group of UVs with different unique IDs.
- **Value**: this constraint type is ensuring that some of the agent class values are limited to certain threshold. For example, the performance value of any UV agent is within the 0 to 100 range.
- **Pre-condition**: this constraint type guarantee the state consistency of an agent instance before triggering the next state. For example, a UV agent can only receive a new task if its current status is 'Idle'. This ensures that an agent must complete its current task or be in a standby state before being assigned a new task, preventing overloading or task conflicts.
- **Post-condition**: this constraint type mandates the new state of an agent instance after moving from old state. For example, after a UV agent has received a new task, its status must be updated to be 'Active'. This reflects that the agent is currently engaged with a task and helps in the accurate tracking and management of the agent's workload.

### 4.3.2 Communication Constraints

OCL, while adept at specifying constraints on the static aspects and behaviors of classes within a UML model, is not inherently designed to manage or constrain communication among the classes themselves. Its limitations become especially pronounced when addressing our case-study requirements to achieve a mission in MAS, where interaction and communication among agents are fundamental (Sadik & Urban, 2018). Multi-agent frameworks like JADE and PADE offer a robust and dynamic ontology language that is called FIPA-ontology communication language (Foundation for Intelligent Physical Agents, 2023). FIPA offers a comprehensive set of interaction protocols, that can achieve intricate patterns of interaction, negotiation, and knowledge exchange among diverse agents, which is a capability OCL doesn't naturally extend.

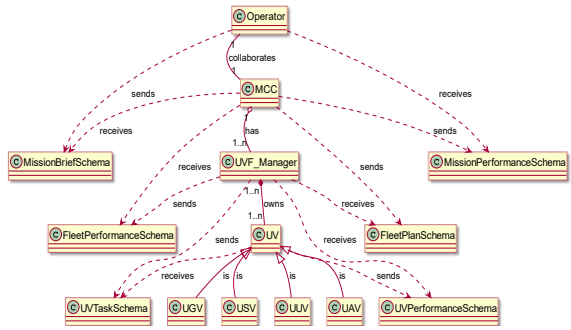

Figure 7: Case-study FIPA-ontology model.

FIPA-ontology is standard that defines a set of fixed schemas that together form the MAS communication model as show in Figure 7. The first set of schemas in our model are the message communication schemas. These set of communication schemas are listed as follows:

- **Mission-Brief**: holds information regarding the mission-brief, containing attributes like mission-ID, description, and status.
- **Fleet-Plan**: contains attributes that provide details about the fleet-plan, such as plan-ID, description, and status.
- **UV-Task**: includes attributes like task-ID, description, and status.
- **UV-Performance**: includes attributes like UV-performance-ID and performance-metric.
- **Fleet-Performance**: includes attributes like Fleet-Performance-ID and performance-metric.
- **Mission-Performance**: includes attributes like mission-performance-ID and performance-metric.

The second set of schemas are the predicates, which depict relationships between agent classes including the communication schemas:

- (**agent-x**) <**is-a**> (**agent-y**): expresses the inheritance relationship between agents, e.g., UAV is a type of UV.
- (**agent-x**) <**has-a**> (**agent-y**): represents the composition relationship between agent, e.g., MCC has a UVF-manager.
- (**agent-x**) <**owns**> (**agent-y**): expresses the aggregation relationship, e.g., the UVF-manager owns multiple UVs.
- (**agent-x**) <**collaborates**> (**agent-y**): defines the collaboration between agents, e.g., the operator collaborates with the MCC.

The third set of schemas are the actions. Actions stand for operations that can be performed by an agent specially in our model on a message schema:

- **Send** (**schema-x**): represents the action of transmitting data, e.g., the operator agent uses this action to send a mission-brief to the MCC.
- **Receive** (**schema-x**): represents the action of receiving data, e.g., the MCC agent uses this action to receive the mission-brief from the operator.

### 4.3.3 Other Constraints

Other technical requirements must be considered in the constraints layer as well. Examples of these constraints can be the auto-generated code quality, code privacy, cybersecurity, etc. One way to formalize these constraints is by using the OCL. For instance, to ensure a consistent indentation we used:

```
self.leadingSpaces.mod(spacePerIndent) = 0.
```

Other OCL constraints that we have considered within this model to regulate the maximum line length, whitespace, function length, and import Statements. Therefore, applying these code quality constraints is ensuring that the auto-generated code is functionally accurate, readable, maintainable, and clear.

Furthermore, formal constraints in this layer can be added to regulate other important aspects of the autogenerated code such as data privacy and cybersecurity. However, in order not to diverse from the main paper topic, we will prefer to discuss that in the future work section.

## 5 CODE EVALUATION

Due to our proposed MDD approach, after modeling the system in a textual formal format such as PlantUML, we give this model as an input to GPT-4 prompt, then we use the output code. In our case study, few bugs have been produced by GPT-4, however by fixing these bugs, the output code was capable of being deployed. As our focus in this study is not the auto-generated code correctness rather than its completeness. We focused on our evaluation on exploring and analyzing the autogenerated code structure and behavior, rather than discovering the type and number of generated code bugs. For this reason, we conducted two different experiments. First Experiment targets the exploring of the autogenerated code behavior, while the second experiment aims to analysis the autogenerated code structure and complexity.

## 5.1 Experiment 1: Behavioural Dynamic analysis

In the first experiment, we orchestrated the generation of two distinct deployments. The first, written in Java, is tailored to run on JADE platform, while the second, crafted in Python, is designated to be implemented in PADE. The goal of the experiment is to compare the code behaviour that run on JADE against the code is running on PADE, to ensure the consistency of the system dynamic regardless the deployment execution language. Accordingly, we observed the agent interaction behavior on JADE vs PADE framework. We found that the agents' behavior as captured by JADE Sniffer tool aligns with the plotted sequence diagram from PADE agent interaction, as shown in Figure 8. Both JADE and PADE deployment have three UV instances, which are UAV, UGV, and USV.

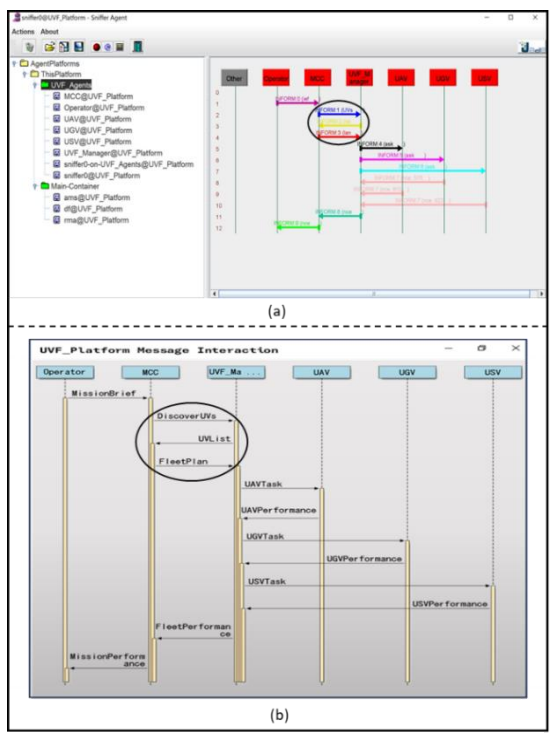

Figure 8: JADE vs PADE Sequence diagram.

In both sequence diagrams in Figure 8, we see that the depicted process commences with the Operator transmitting a mission-brief to the MCC. On receiving this, the MCC solicits the UVF-manager to identify available UVs. Upon obtaining a list of accessible UVs, the MCC devises a fleet-plan and conveys it to the UVF-manager. After this, the UVF-manager dispatches specific tasks to the UAV, UGV, and USV. Each UV, upon task completion, relays performance data to the UVF-manager. Collating this data, the UVF-manager formulates a comprehensive fleet-performance metric, which is relayed back to the MCC. The MCC, in turn, evaluates this metric in congruence with the mission objectives, compiling a definitive mission-performance report. This report, the culmination of the entire operation, is ultimately returned to the Operator.

Two important remarks have been noticed from comparing these two sequence diagrams in Figure 8 with the original case-study activity diagram in Figure 4. First, we noticed that ChatGPT has enhanced the interaction by adding new behaviours to MCC agent and UVF-manager agent. This new behaviour can be seen when the MCC is sending DiscoverUVs message to the UVF-Manger agent and waiting the UVList before forming a FleetPlan, as logically the MCC needed to know what the available UVs resources are before planning them based on the mission-brief. This new interaction behaviour was not explicitly mentioned in the case-study activity diagram. The second remark that the timing of interaction between the MCC and the UVs differ in JADE and PADE, most probably due to the difference in the state machine of each UV instance. This is a good indication that these UV state machine can emulate the operation of the agents.

## 5.2 Experiment 2: Structural Complexity assessment

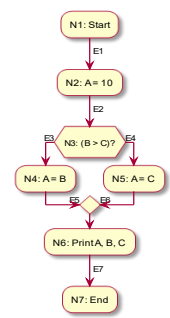

Figure 9: code control-flow graph example.

The second experiment focuses on exploring the autogenerated code structure and complexity. Therefore, in this experiment we used the cyclomatic complexity metric, to measure and analysis the autogenerated code complexity. Cyclomatic complexity quantifies the code complexity by counting the number of linearly independent paths through its source code.

Calculated using the control-flow graph of the code as the one shown in Figure 9. In the control-flow graph example shown in Figure 9, the cyclomatic complexity (M) can be calculated from the formula:

$$M = E - N + 2P \qquad (1)$$

Where:
- E is the number of edges in the flow graph
- N is the number of nodes
- P is the number of graph separate branches

Thus, M in this case equals 3.

Accordingly, M can be used to assess the difficulty of code testing, maintenance, understanding, refactoring, performance, reliability, and documentation, where the following values are considered in the assessment:
- M = 1-10: law risk
- M = 11-20: moderate risk
- M = 21-50: high risk; needs to be reviewed and perhaps split into smaller modules
- M > 50: sever risk; necessary refactoring is required

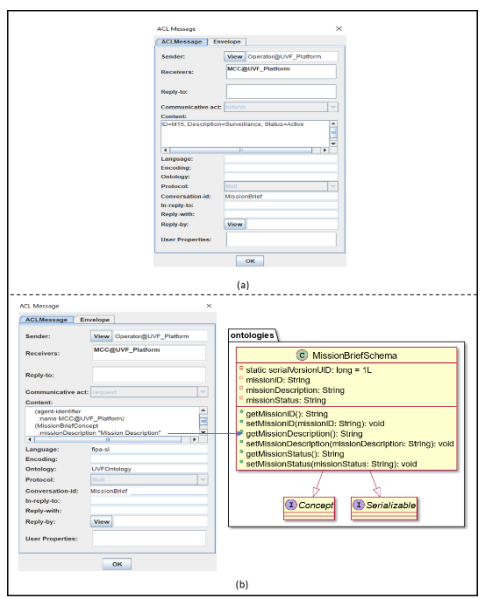

Figure 10: String based communication vs schema-based communication.

As in our MDD approach, we emphasized the effect of adding the formal constraints on generating a deployed code, our interest in this experiment is to understand the influence of the constraints layer on the autogenerated code. Therefore, in the experiment we autogenerated two distinct deployments, that differ in the level of constraints involved in their models. The first model implements only the OCL constraints, while the second model add the FIPA-ontology to the model. In case of using OCL constraints only, the agents are communicating via string-based message as shown in the agent communication message in Figure 10-a, while using OCL along with FIPA-ontology constraints resulted in agents that communicate via schema-based message as shown in Figure 10-b.

Table 1: Cyclomatic Complexity of auto -generated code model with OCL constraints only.

| Agent class | Operator | MCC | UVF-Manager | UV | Model |
|---|---|---|---|---|---|
| Edges (E) | 8 | 15 | 16 | 8 | |
| Nodes (N) | 8 | 13 | 14 | 8 | |
| Branches (P) | 1 | 1 | 1 | 1 | |
| Complexity (M) | 2 | 4 | 4 | 2 | 12 |

Table 2: Cyclomatic Complexity of auto -generated code with OCL and Ontology constraints.

| Agent class | Operator | MCC | UVF-Manager | UV | Model |
|---|---|---|---|---|---|
| Edges (E) | 12 | 22 | 23 | 12 | |
| Nodes (N) | 11 | 19 | 19 | 11 | |
| Branches (P) | 1 | 1 | 1 | 1 | |
| Complexity (M) | 3 | 5 | 6 | 3 | 17 |

After generating the two distinct deployments, we transformed the agent classes into control flow diagrams to calculate their M, as shown in Table 1 and Table 2. By comparing M values of the auto-generated code in Table 1 and Table 2, we will find that the code complexity is slightly increasing by adding the FIPA-ontology constraints in the second the deployment. however, the complexity of all the agent classes in both deployments is still locating under the law risk category. This means that the auto-generated structure is adequate and does not need any further refactoring. Furthermore, the highest M value belongs to the UVF-manger in the second deployment, where we considered both the OCL and the FIPA-ontology constraints. This value equals to 6, which means that there is a still a large risk marge that allows us to add further constraints in our model without negatively influencing the complexity of the autogenerated code.

## 6 DISCUSSION, CONCLUSION, AND FUTURE WORK

In our research, we highlighted the difficulties in auto-generating deployable code from natural language using LLMs like ChatGPT, primarily due to language ambiguity. To address this, we employed formal modelling languages, such as UML, for better interpretation by ChatGPT. We found that current UML code generation practices don't fully exploit LLMs, revealing a gap in agility within the MDD process. To enhance this agility, we introduced "constraints" into UML models, adding semantic depth to ensure accurate code generation. These constraints improve various software aspects, such as structure, and communication.

In our case study, we showcased our proposed MDD approach by modelling a multi-agent of UVF. We used class diagrams to outline agents, while activity and state diagrams captured their interactions and internal behaviours. Detailed constraints were provided using the Object Constraint Language (OCL) for structure, and FIPA-ontology for agent communication. This model then served as a foundation for auto-generating code in both Java and Python using GPT-4, chosen for its advanced reasoning over GPT-3.5. The effectiveness of our MDD approach relies on the LLM's ability to accurately understand the model's constraints, ensuring code generation remains true to our design.

In the first evaluation experiment, we examined the behaviour of auto-generated code within simulation environments: Java's JADE and Python's PADE frameworks. Both deployments effectively captured the intended agent interactions, though there were minor sequence variations between them. Remarkably, GPT-4 not only adhered to the specified agent logic but also enriched it by introducing two new behaviours in the MCC agent's communication sequence. This addition highlighted the power of communication constraints in guiding GPT-4 and its enhanced comprehension of agent interactions. While these improvements were impressive, they underscored a need for meticulous code review. Despite GPT-4's advancements, ensuring that the generated code remains consistent with design intentions is crucial to prevent unexpected behaviours.

In the second experiment, we examined the structural of the auto-generated code, specifically by assessing its cyclomatic complexity. In this experiment, we created two separate deployments. The first deployment code resulted from a model that involves only OCL constraints, while the second deployment coded resulted from a model that involves both OCL and FIPA-Ontology constraints. Our analysis revealed the intriguing remake that integrating FIPA-ontology constraints didn't dramatically augment the complexity of the auto-generated code. This suggests that these constraints provide meaningful semantics without unduly complicating the resultant codebase. Furthermore, the analysis also hinted at a notable latitude in our approach. There appears to be a reasonable buffer allowing for the inclusion of additional constraints to the model in future iterations without triggering an immediate need for a code refactor. This is indicative of the robustness and scalability inherent in our MDD approach.

In our exploration into integrating the advantages of LLMs into MDD, we've identified that using formal modelling languages can significantly bridge the gap between the challenges of natural language ambiguities and the precision of code generation. The incorporation of meta-modeling constraints not only refines the code generation process but also provides insights into its structural complexity, ensuring a more informed and resilient codebase. Combined, these advancements hint at a transformative path to achieving the elusive agility in current MDD practices. As the world of software development evolves, this seamless interplay between structured modeling, advanced LLM reasoning, and structural complexity assessments will be paramount in crafting agile, efficient, and robust software solutions.

In upcoming research, we plan to assess the correctness of the auto-generated code by quantifying the bugs present and pinpointing if certain defects consistently relate to the model. Given the influential role of constraints in refining the auto-generated code, we intend to incorporate new privacy and cybersecurity constraints and subsequently analyse the characteristics of the resultant code. It's also essential to compare our methodology with current MDD frameworks, evaluating factors like efficiency, accuracy, and reliability in various contexts. Through comprehensive enhancement and evaluation, we aim to pave the way for broader industry adoption.

## REFERENCES

Ahmad, A., Waseem, M., Liang, P., Fehmideh, M., Aktar, M. S., & Mikkonen, T. (2023). Towards Human-Bot Collaborative Software Architecting with ChatGPT (arXiv:2302.14600).

Belghiat, A., & Bourahla, M. (2012). From UML Class Diagrams to OWL Ontologies: A Graph Transformation Based Approach. ICWIT, 330–335.

Brulin, S., & Olhofer, M. (2023). Bi-level Network Design for UAM Vertiport Allocation Using Activity- Based Transport Simulations.

Cabot, J., & Gogolla, M. (2012). Object constraint language (OCL): A definitive guide. In International school on formal methods for the design of computer, communication and software systems (pp. 58–90). Springer.

Cámara, J., Troya, J., Burgueño, L., & Vallecillo, A. (2023). On the assessment of generative AI in modeling tasks: An experience report with ChatGPT and UML. Software and Systems Modeling, 22(3), 781–793.

Chen, M., Tworek, J., Jun, H., Yuan, Q., Pinto, H. P. de O., Kaplan, J., Edwards, H., Burda, Y., Joseph, N., Brockman, G., Ray, A., Puri, R., Krueger, G., Petrov, M., Khlaaf, H., Sastry, G., Mishkin, P., Chan, B., Gray, S., … Zaremba, W. (2021). Evaluating Large Language Models Trained on Code (arXiv:2107.03374).

David, I., Latifaj, M., Pietron, J., Zhang, W., Ciccozzi, F., Malavolta, I., Raschke, A., Steghöfer, J.-P., & Hebig, R. (2023). Blended modeling in commercial and open-source model-driven software engineering tools: A systematic study. Software and Systems Modeling, 22(1), 415–447. https://doi.org/10.1007/s10270-022-01010-3

Dong, Y., Jiang, X., Jin, Z., & Li, G. (2023). Self-collaboration Code Generation via ChatGPT

Fadjukoff, L., & Tolvanen, J.-P. (2022). Comparing the Effort of Developing Enterprise Applications with Programming and with Domain-Specific Modeling.

Feltus, C., Grandry, E., Kupper, T., & Colin, J.-N. (2017). Model-driven Approach for Privacy Management in Business Ecosystem: Proceedings of the 5th International Conference on Model-Driven Engineering and Software Development, 392–400.

FIPA, S. (2000). FIPA ontology service specification. Citeseer.

Foundation for Intelligent Physical Agents. (2023). http://www.fipa.org/

Hailpern, B., & Tarr, P. (2006). Model-driven development: The good, the bad, and the ugly. IBM Systems Journal, 45(3), 451–461.

Kapferer, S., & Zimmermann, O. (2020). Domain-specific Language and Tools for Strategic Domain-driven Design, Context Mapping and Bounded Context Modeling: Proceedings of the 8th International Conference on Model-Driven Engineering and Software Development, 299–306.

Kelly, S., & Tolvanen, J.-P. (2008). Domain-Specific Modeling: Enabling Full Code Generation. John Wiley & Sons.

Liu, C., Bao, X., Zhang, H., Zhang, N., Hu, H., Zhang, X., & Yan, M. (2023). Improving ChatGPT Prompt for Code Generation (arXiv:2305.08360).

OMG, O. I. (2006). Object management group. Needham, MA, USA, 2(2).

Perez-Martinez, J. E., & Sierra-Alonso, A. (2004). UML 1.4 versus UML 2.0 as languages to describe software architectures. Lecture Notes in Computer Science (Including Subseries Lecture Notes in Artificial Intelligence and Lecture Notes in Bioinformatics), 3047.

Petrovic, N., & Al-Azzoni, I. (2023). AUTOMATED APPROACH TO MODEL-DRIVEN ENGINEERING LEVERAGING CHATGPT AND ECORE.

Sadik, A. R., Bolder, B., & Subasic, P. (2023). A self-adaptive system of systems architecture to enable its ad-hoc scalability: Unmanned Vehicle Fleet - Mission Control Center Case study. Proceedings of the 2023 7th International Conference on Intelligent Systems, Metaheuristics & Swarm Intelligence, 111–118.

Sadik, A. R., Ceravola, A., Joublin, F., & Patra, J. (2023). Analysis of ChatGPT on Source Code (arXiv:2306.00597).

Sadik, A. R., & Goerick, C. (2021). Multi-Robot System Architecture Design in SysML and BPMN. Advances in Science, Technology and Engineering Systems Journal, 6(4). https://doi.org/10.25046/aj060421

Sadik, A. R., & Urban, B. (2018). CPROSA-Holarchy: An Enhanced PROSA Model to Enable Worker–Cobot Agile Manufacturing. International Journal of Mechanical Engineering and Robotics Research, 7(3).

Sarkisian, A., Vasylkiv, Y., & Gomez, R. (2022). System Architecture Supporting Crowdsourcing of Contents for Robot Storytelling Application.

Sharaf, M., Abusair, M., Eleiwi, R., Shana'a, Y., Saleh, I., & Muccini, H. (2019). Modeling and Code Generation Framework for IoT (pp. 99–115).

Siricharoen, W. (2009). Ontology Modeling and Object Modeling in Software Engineering. International Journal of Software Engineering and Its Applications, 3.